\documentclass[a4paper,useAMS,usenatbib]{mn2e}
\usepackage[totalwidth=480pt, totalheight=680pt]{geometry}
\usepackage{graphicx}

\def\nup{$\nu_{\rm peak}^S$}
\def\gr{{$\gamma$-ray}}
\newcommand{\fermi}{{\it Fermi} }

\newcommand{\lsim}{{\lower.5ex\hbox{$\; \buildrel < \over \sim \;$}}}
\newcommand{\gsim}{{\lower.5ex\hbox{$\; \buildrel > \over \sim \;$}}}
\newdimen\digitwidth
\setbox0=\hbox{2}
\digitwidth=\wd0
\catcode `#=\active
\def#{\kern\digitwidth}

\title[High power -- high synchrotron peak blazars] {The discovery of high
  power -- high synchrotron peak blazars}
\author[P. Padovani, P. Giommi, and A. Rau]{P. Padovani$^{1}$\thanks{E-mail:
ppadovan@eso.org}, P. Giommi$^{2}$, and A. Rau$^{3}$\\
$^{1}$European Southern Observatory, Karl-Schwarzschild-Str. 2,
                   D-85748 Garching bei M\"unchen, Germany\\
$^{2}$ASI Science Data Center, c/o ESRIN, via G. Galilei, I-00044 Frascati,
                   Italy \\
$^{3}$Max-Planck-Institut f\"ur Extraterrestrische Physik,
Giessenbachstr. 1, D-85748 Garching bei M\"unchen, Germany}
\begin{document}
\newdimen\digitwidth
\setbox0=\hbox{2}
\digitwidth=\wd0
\catcode `#=\active
\def#{\kern\digitwidth}

\date{Accepted ... Received ...; in original form ...}

\pagerange{\pageref{firstpage}--\pageref{lastpage}} \pubyear{2012}

\maketitle

\label{firstpage}

\begin{abstract}
We study the quasi-simultaneous near-IR, optical, UV, and X-ray photometry
of eleven $\gamma$-ray selected blazars for which redshift estimates larger
than 1.2 have been recently provided. Four of these objects turn out to be
high-power blazars with the peak of their synchrotron emission between
$\sim 3\times 10^{15}$ and $\sim 10^{16}$ Hz, and therefore of a kind
predicted to exist but never seen before. This discovery has important
implications for our understanding of physical processes in blazars,
including the so-called ``blazar sequence'', and might also help
constraining the extragalactic background light through $\gamma$-ray
absorption since two sources are strongly detected even in the 10$-$100 GeV
\fermi-LAT band. Based on our previous work and their high powers,
these sources are very likely high-redshift flat-spectrum radio quasars
with their emission lines swamped by the non-thermal continuum.
\end{abstract}

\begin{keywords}
  BL Lacertae objects: general --- radiation mechanisms: non-thermal ---
  gamma-rays: observations --- radio continuum: galaxies --- X-rays:
  galaxies
\end{keywords}

\section{Introduction}\label{intro}

Blazars are radio loud active galactic nuclei (AGN) with their jets
pointing towards the observer \citep[see e.g. ][]{bla78,UP95}.
Observationally, they come in two main flavours with very different optical
spectra.  Namely, Flat Spectrum Radio Quasars (FSRQs) have strong, broad
emission lines, just like radio quiet quasars, while BL Lacertae objects
(BL Lacs) show at most weak emission lines, sometimes display absorption
features, and can also be completely featureless.

Blazars emit variable, non-thermal radiation across the whole
electromagnetic spectrum, which includes two components forming two broad
humps in a $\nu - \nu f_{\nu}$ representation. The low-energy one is
attributed to synchrotron radiation and the high-energy one is usually
thought to be due to inverse Compton radiation \citep[e.g.][]{abdosed}. The
peak of the synchrotron emission (\nup) can occur over a wide range of
frequencies, from about $\sim 3 \times 10^{12}$ Hz to over $10^{18}$ Hz,
reflecting the maximum energy at which particles can be accelerated
\citep[e.g.][]{GiommiPlanck}. Blazars with rest-frame \nup~$< 10^{14}$~Hz
are called Low Synchrotron Peaked (LSP) sources, while those with
$10^{14}$~Hz $<$ \nup~$< 10^{15}$~Hz and \nup~ $> 10^{15}$~Hz are called
Intermediate and High Synchrotron Peaked (ISP and HSP) sources,
respectively \citep{abdosed}. These definitions expand on the original
division of BL Lacs into LBL and HBL sources first introduced by
\cite{padgio95}.

The location of the synchrotron peak implies different mechanisms for the
X-ray emission, namely inverse Compton in LSPs, which have hard energy
index ($\alpha_{\rm x} \lsim 1$) spectra, and an extension of the
synchrotron emission responsible for the lower energy continuum in HSPs,
which typically display steeper ($\alpha_{\rm x} \approx 1.5$) X-ray
spectra\footnote{Unless \nup~is in the X-ray band, where one samples the
  top of the synchrotron hump and the X-ray spectrum is also relatively
  flat \citep{padgio96}.} \citep[e.g.][]{padgio96,wol98,pad04}.

The distribution of the synchrotron peak frequency is extremely different
for the two blazar classes. While the rest-frame \nup~distribution
of FSRQs is strongly peaked at low frequencies ($\langle$\nup
$\rangle=10^{13.1\pm0.1}$ Hz) and never reaches high values (\nup $\lsim
10^{14.5}$ Hz) \citep{GiommiPlanck}, the \nup~distribution of BL Lacs is
shifted to higher values by at least one order of magnitude and can reach
\nup~as high as $\gsim 10^{18}$ Hz. In other words, while BL Lacs can be of
the LSP and HSP type, FSRQs appear to be solely LSPs.

The origin of this difference has been discussed at length in the
literature. In a series of papers, Padovani, Giommi, and collaborators,
looked for HSP FSRQs using multi-wavelength catalogues and the deep X-ray
radio blazar survey \citep[DXRBS; e.g.][]{pad03} with only limited
success. In fact, while FSRQs with \nup~possibly larger than previously
known were discovered, follow-up X-ray observations showed in most cases a
relatively flat X-ray spectrum. This, together with their overall spectral
energy distributions (SEDs), was suggestive of inverse Compton emission
\citep[e.g.][]{pad02,lan08} and did not support a possible HSP
classification. The only exception is RGB J1629+4008, with \nup~$\sim 2
\times 10^{16}$~Hz and steep X-ray spectrum \citep{pad02}. ROXA
J081009.9+384757.047 was also suggested to be an HSP \citep{gio07} but new
WISE data \citep{wri10} indicate that the near-IR and optical flux can be
attributed to accretion (the big ``blue bump'') and that the peak of the
non-thermal synchrotron radiation is located in the far infrared, making
this object a typical LSP.

The purported non-existence of HSP FSRQs implied also the lack of
high-power HSP blazars (since FSRQs are more powerful than BL Lacs), which
was also hinted at by a suggested anti-correlation between bolometric
luminosity and \nup~\citep[the so-called ``blazar sequence";][]{fossati98}. 
This has been interpreted by some researchers as an {\it intrinsic}
physical difference between LSPs and HSPs. Since HSPs are typically
characterized by a low intrinsic power and external radiation field, given
their very weak or absent emission lines, cooling was thought to be less
dramatic than in LSPs allowing particles to reach energies high enough to
produce synchrotron emission well into the X-ray band \citep{ghis98}. In
this respect, RGB 1629+4008, with its relatively low radio power, $P_{\rm 5
  GHz} \sim 4 \times 10^{24}$ W Hz$^{-1}$, was regarded as fitting on the
sequence as well \citep{pad02}.

\cite{gio12} have recently presented a new scenario, where blazars are
classified as FSRQs, BL Lacs, LSPs, or HSPs according to a varying mix of
the Doppler boosted radiation from the jet, emission from the accretion
disk, the broad line region, and the host galaxy. This hypothesis predicts
the existence of high power -- high \nup~blazars, which however would be
very hard to identify because of their featureless optical spectra and,
therefore, lack of redshift. This is because when both \nup~and radio power
are large the dilution by the non-thermal continuum becomes extreme and all
optical features are washed away.

Very recent results provide a simple way to test this prediction and look
for the existence of previously unidentified high power HSPs. \cite{rau12}
have presented redshift constraints for 103 blazars from the {\it Fermi}
2LAC catalogue \citep{fermi2lac} by fitting SED templates to their
UV-to-near-IR multi-band photometry obtained quasi-simultaneously with {\it
  Swift}/UVOT and GROND \citep{greiner}. In fact, attenuation due to
neutral hydrogen along the line of sight at the Lyman limit allows an
accurate estimate of the redshift of the absorber and thereby provides a
reliable lower limit to the blazar redshift. Eleven of their objects have
$z_{\rm phot} > 1.2$ and for eight such sources these are the first
reliable redshift measurements.  Since the spectrum of most of these
sources is featureless and high redshift typically means high power, this
is an ideal sample to look for high power HSPs with their lines completely
swamped by the non-thermal continuum, which is the purpose of this work.

Throughout this paper we use a $\Lambda$CDM cosmology with $H_0 = 70$ km
s$^{-1}$ Mpc$^{-1}$, $\Omega_m = 0.27$ and $\Omega_\Lambda = 0.73$
\citep{kom11}.

\section{SEDs of the newly identified high redshift blazars}\label{sed}

\begin{figure}
\centering
\includegraphics[width=8.5cm]{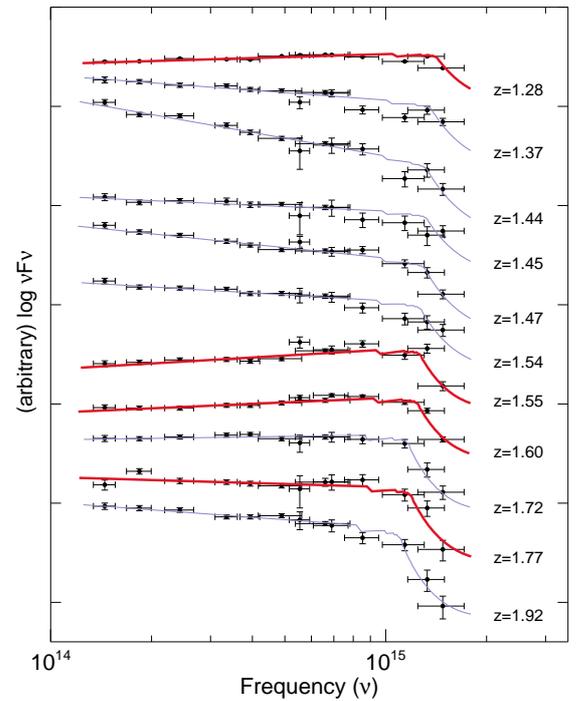}
\caption{The near-IR through UV SEDs for the eleven blazars with $z_{\rm
    phot} > 1.2$ from \protect\cite{rau12} in a $\nu - \nu f_{\nu}$ format. The
  lines are the best fit SEDs with red, thick lines indicating our
  HSPs. From bottom to top: CRATES J0402$-$2615, CRATES J1312$-$2156, OM
  235, CRATES J0630$-$2406, SUMSS J053748-571828, PKS 0600$-$749, PKS
  0332$-$403, CLASS J2352+1749, PKS0047+023, PKS 0055$-$328, RX
  J0035.2+1515.}
\label{fig:nufnu_all} 
\end{figure}

Fig. \ref{fig:nufnu_all} plots the near-IR through UV SEDs of all eleven
blazars with $z_{\rm phot} > 1.2$ from \cite{rau12} in a $\nu - \nu
f_{\nu}$ representation. Taking into account the obvious decrease in flux
at high frequencies due to neutral hydrogen absorption, it is already clear
from these SEDs covering only about one decade in frequency that a few of
the sources do appear to have \nup~$\gsim 10^{15}$ Hz, which qualifies them
as HSPs.  We have then assembled the SEDs of all eleven sources over the
whole electromagnetic spectrum by using the SED builder tool available at
the ASDC \citep[tools.asdc.asi.it/SED,][]{SEDtool}. As mentioned in the
Introduction, a key requirement for an HSP classification is a steep X-ray
spectrum (for \nup~$\lsim 10^{17}$ Hz). Since the {\it Swift}/X-ray
Telescope (XRT) observes contemporaneously with {\it Swift}/UVOT, our
sources have {\it quasi-simultaneous coverage} from the GROND bands into
the X-ray regime, i.e. between $10^{14}$ and $10^{18}$ Hz. This is an ideal
situation to determine \nup~for HSPs.

\begin{table*}
\caption{Observational parameters}
\begin{tabular}{lllcllcl}
 Name & {\it Fermi} Name & Optical Coordinates$^a$ & $z_{\rm phot}^a$ & $f_{\rm 1.4 GHz}$  & ${r^{\prime}}^a$ & $f_{2-10 keV}$  & $f_{1-100 GeV}$\\
              &       &     &    &  mJy  & AB  & erg/cm$^{2}$/s &   $10^{-10}$ ph/cm$^2$/s\\
 \hline
RX J0035.2+1515 & 2FGLJ0035.2+1515 & 00 35 14.7 +15 15 04.2 & $1.28^{+0.14}_{-0.17}$ & #$18.8$ & $17.10$ & $3.3~10^{-13}$ & $10.8$ \\
SUMSS J053748$-$571828  & 2FGLJ0537.7$-$5716 & 05 37 49.0 $-$57 18 30.4 & $1.55^{+0.09}_{-0.13}$ & #$99.8^b$ & $17.48$ &  $5.5~10^{-13}$ & #$3.9$\\
CRATES J0630$-$2406 & 2FGLJ0630.9$-$2406 & 06 30 59.5 $-$24 06 46.2 & $1.60^{+0.10}_{-0.05}$ & $105.8$ & $15.90$ & $10.7~10^{-13}$ & $33.8$\\
CRATES J1312$-$2156 &  2FGLJ1312.4$-$2157 & 13 12 31.6 $-$21 56 23.5 & $1.77^{+0.09}_{-0.11}$ &  349.8 & 16.56 &  $5.1~10^{-13}$ &$21.7$\\
\hline
\multicolumn{5}{l}{\footnotesize $^a$From \cite{rau12}}\\
\multicolumn{5}{l}{\footnotesize $^b$Radio flux density at 843 MHz}\\
\end{tabular}
\label{tab:obs}
\end{table*}

It turns out that four sources are HSPs, whose main observational
parameters are given in Table \ref{tab:obs}. We determined rest-frame
\nup~values by fitting a third degree polynomial\footnote{This represents
  the simplest way to estimate \nup~in a straightforward and
  model-independent way.} to the radio data, the GROND and UVOT data below
the Lyman limit, and the quasi-simultaneous {\it Swift}/XRT data.  We also
fitted the X-ray data (using the XSPEC12.0 package) adopting a simple power
law spectral model with the low energy absorption fixed to the Galactic
value in the direction of the source \citep{di90}. In two cases, given the
relatively short exposure times, we used the XIMAGE package
\citep{giommiximage} V4.5\footnote{XIMAGE is part of the HEASARC Xanadu
  standard software package for multi-mission X-ray astronomy.} to estimate
the source count rates in the soft ($0.5-2.0$ keV) and hard ($2.5-10$ keV)
X-ray energy bands and derive an estimate of the spectral slope. In the
case of CRATES J1312$-$2156, however, we could only get a $3\sigma$ upper
limit at higher energies, which translates into a lower limit to the slope.

\begin{table*}
\caption{Inferred parameters}
\begin{tabular}{lccllc}
 Name & log \nup~& log $L_{\rm peak}$ & log $P_{\rm 1.4 GHz}$ & $\alpha_{\rm x}$  & $\alpha_{\gamma}$$^a$ \\
             &      Hz  (rest-frame) &   erg s$^{-1}$   &   W Hz$^{-1}$    & (0.5 -- 10 keV)   &   (100 MeV -- 100 GeV) \\
\hline
RX J0035.2+1515 & 15.8  & 46.6 & 26.2 & $1.9^{+0.1}_{-0.1}$ & $0.61\pm0.11$ \\
SUMSS J053748$-$571828 & 15.5  & 46.6 & 27.1$^b$ & $1.6^{+0.5}_{-0.3}$$^c$ & $0.73\pm0.21$\\
CRATES J0630$-$2406 & 16.0  & 47.1 & 27.1 & $1.9^{+0.2}_{-0.2}$ & $0.79\pm0.06$\\
CRATES J1312$-$2156 & 15.6  & 46.9 & 27.7 & $>0.5$$^c$ & $1.02\pm0.07$ \\
\hline
\multicolumn{5}{l}{\footnotesize $^a$From \cite{fermi2fgl}}\\
\multicolumn{5}{l}{\footnotesize $^b$Extrapolated from 843 MHz assuming $\alpha_{\rm r} = 0$}\\
\multicolumn{5}{l}{\footnotesize $^c$Derived from the count rates in the $0.5 - 2.0$ and $2.5 -10$ keV bands}\\
\end{tabular}
\label{tab:inf}
\end{table*}

\begin{figure}
\centering
\includegraphics[height=8.0cm,angle=-90]{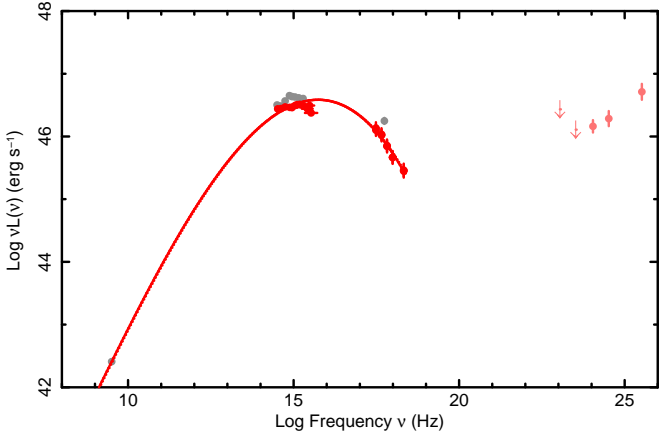}
\hspace{0.3cm}
\vspace{1.5cm}
\caption{The rest-frame SED of RX J0035.2+1515. Red points represent the
  quasi-simultaneous GROND and {\it Swift} data. The light red \gr\ points
  are from the second \fermi-LAT catalog \citep{fermi2fgl}. Gray points
  are non-simultaneous archival data. The solid line is a third degree
  polynomial fit to the radio points, simultaneous GROND and {\it Swift} UVOT
  data not affected by UV absorption, and XRT data.}
\label{fig:0035} 
\end{figure}

\begin{figure}
\centering
\includegraphics[height=8.0cm,angle=-90]{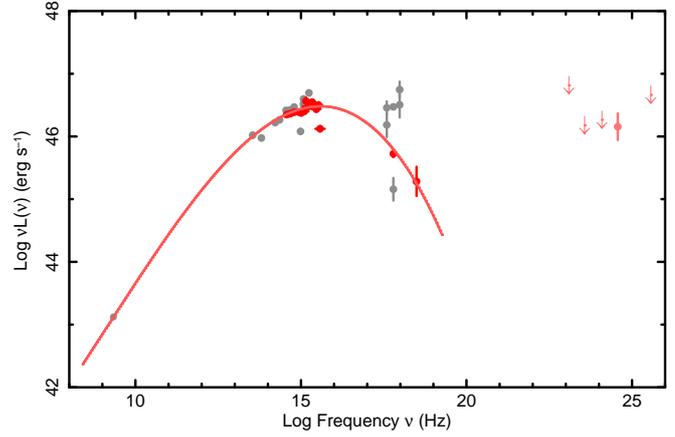}
\hspace{0.3cm}
\vspace{1.5cm}
\caption{The rest-frame SED of SUMSS J053748$-$571828. Data points and solid
  line are as in Fig. \ref{fig:0035}.}
\label{fig:0537} 
\end{figure}

\begin{figure}
\centering
\includegraphics[height=8.0cm,angle=-90]{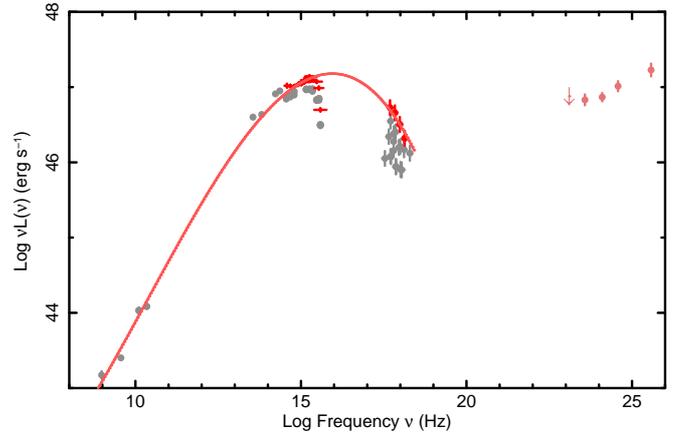}
\hspace{0.3cm}
\vspace{1.5cm}
\caption{The rest-frame SED of CRATES J0630$-$2406. Data points and solid
  line are as in Fig. \ref{fig:0035}.}
\label{fig:0630} 
\end{figure}

\begin{figure}
\centering
\includegraphics[height=8.0cm,angle=-90]{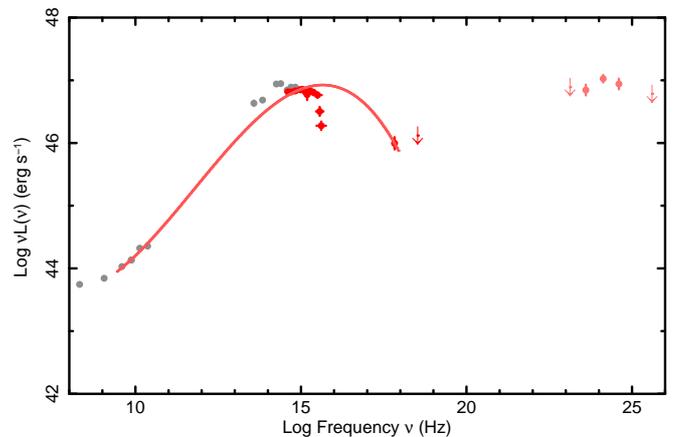}
\hspace{0.3cm}
\vspace{1.5cm}
\caption{The rest-frame SED of CRATES J1312$-$2156. Data points and solid
  line are as in Fig. \ref{fig:0035}.}
\label{fig:1312} 
\end{figure}

The SEDs for the HSPs are shown in Figs. \ref{fig:0035} -- \ref{fig:1312}
while Table \ref{tab:inf} gives \nup, $L_{\rm peak}$, $P_{\rm 1.4 GHz}$,
$\alpha_{\rm x}$, and $\alpha_{\gamma}$. All four sources are characterized
by \nup~$> 3 \times 10^{15}$ Hz and a rising $\gamma$-ray spectrum
($\alpha_{\gamma} \le 1$). Three of the sources display also a very steep
X-ray spectrum ($\alpha_{\rm x} \ge 1.6$), while for the fourth one we can
only say that $\alpha_{\rm x} > 0.5$. $L_{\rm peak}$ and $P_{\rm
  1.4 GHz}$ are quite high, being $> 4 \times 10^{46}$ erg s$^{-1}$ and $>
2 \times 10^{26}$ W Hz$^{-1}$, i.e. typical of FSRQs. We also note that SUMSS
J053748$-$571828 shows clear evidence of X-ray variability of the type
displayed by HSPs in the X-ray band when this is beyond~\nup. Similarly,
CRATES J0630$-$2406 also shows evidence of X-ray variability, with {\it
  Swift}/XRT data taken at different times still displaying a steep X-ray
spectrum. 

\section{Discussion and Implications}

All of the observational evidence suggests that we have discovered four
blazars of a previously unknown type, that is sources having both high
power ($L_{\rm peak} \approx 6 \times 10^{46}$ erg s$^{-1}$) and high
\nup~($\approx 5 \times 10^{15}$ Hz). These objects have also radio powers
$2-3$ orders of magnitude larger than the only previously known HSP FSRQ
and therefore are firmly into the FSRQ regime \citep[unlike RGB 1629+4008,
  which was near the low-luminosity end of the FSRQ luminosity
  function:][]{pad02}.

\begin{figure}
\centering
\includegraphics[height=7.1cm,angle=-90]{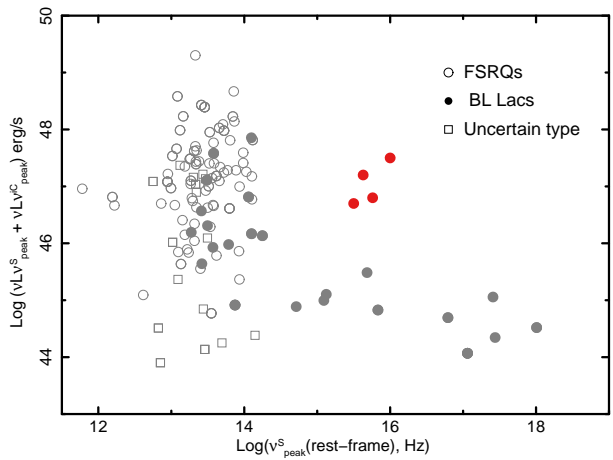}\\
\vspace{1.5cm}
\caption{The bolometric power (represented by the sum of the synchrotron
  and inverse Compton peak powers) of the flux-limited samples of blazars
  of \protect\cite{GiommiPlanck} is plotted against \nup~\protect\citep[see][for
    details]{GiommiPlanck}. The four objects discovered in this paper,
  represented by the red filled points, are clearly located in an area of
  the diagram that was not populated in the past.}
\label{fig:blazseq} 
\end{figure}

To put things into perspective, Fig. \ref{fig:blazseq}, adapted from
\cite{GiommiPlanck}, plots the bolometric power (i.e., the sum of the
synchrotron and inverse Compton peak powers) for blazars selected in the
radio, X-ray and \gr-bands, against \nup. Our HSPs end up in an empty
region of the plot, where no objects were found before. Similar results are
obtained if all 2LAC BL Lacs with redshifts are also included.

Our \nup~estimates are robust, as they are based on quasi-simultaneous
coverage over four decades in frequency and are corroborated in at least
three cases by a steep X-ray spectrum. Contamination by a thermal
(accretion disk) component is ruled out by the featureless spectra of these
sources and also by the fact that the flux of an accretion disk fitted
using the scaling of \cite{gio12} would be well below the observational
data. Our sources also end up in the HBL region of the $\alpha_{\rm
  x}$ -- \nup~plot of \cite{pad04} (their Fig. 5).

These four sources are also the only ones with $\alpha_{\rm \gamma} \le 1$,
as the other seven in the \cite{rau12} sample have $\alpha_{\rm \gamma} >
1$ \citep{fermi2fgl}. While one is borderline with $\alpha_{\rm \gamma}
\sim 1$, three have $\alpha_{\rm \gamma} \le 0.79$. We note that no FSRQ,
LSP, or ISP reach these low values in the Clean Sample of \cite{fermi2lac}.
It is also interesting to note that 4/6 of the remaining objects in the
\cite{rau12} sample (excluding the FSRQ OM 235) have \nup~$> 10^{14}$ Hz
and therefore qualify as ISPs. 

In the radio power -- \nup~plot of \cite{gio12} (their Fig. 11) the newly
discovered HSPs end up in the region of parameter space mostly populated by
sources without redshift, which agrees with the fact that, indeed, these
sources have totally featureless optical spectra. Based on that paper,
  our sources are very likely FSRQs with their emission lines swamped by
  the non-thermal continuum.  Given their large powers, this makes sense
  also on statistical grounds. Indeed, for $L_{\rm bol} > 5 \times 10^{46}$
  erg s$^{-1}$, which is the smallest value for our sources, previously
  known FSRQs outnumber BL Lacs by a factor $> 10$ in
  Fig. \ref{fig:blazseq}.

The discovery of high power HSPs has important implications and consequences
for a variety of issues. We mention here some of the most prominent ones:

\begin{enumerate}

\item our HSPs end up in an empty region of the $L_{\rm peak}$ -- \nup~plot
  of \cite{mey11} (see their Fig. 4). In fact, the closest sources with
  \nup~similar to those of our objects have $L_{\rm peak}$ more than an
  order of magnitude smaller. This implies that the so-called ``blazar
  sequence'' \citep{fossati98,ghis98} is indeed heavily influenced by the
  selection effects described by \cite{gio12};

\item it is well established that a large fraction of BL Lacs \citep[$\sim
  43$\% in BZCAT \cite{mas09} and $> 50 - 60$\% of the BL Lacs in the
  Fermi 1 and 2 year AGN catalogs:][]{fermi1lac,fermi2lac} have no measured
  redshift due to the lack of any detectable feature in their optical
  spectrum, despite the use of 8/10-m class optical telescopes for the
  spectroscopy identification campaign. This paper, together with the
  results of \cite{rau12}, confirms the suggestion by \cite{gio12} that
  these sources are high-redshift blazars, most likely FSRQs with their
  emission lines swamped by the jet. Therefore: a) one should not assume
  for these objects a redshift typical of other BL Lacs but a much larger
  one (typically $z > 1$); b) these sources should be included with the
  FSRQs when studying cosmological evolution and luminosity functions,
  since their exclusion is bound to bias the results;

\item the study of distant very high energy (VHE) emitting AGN is of
  fundamental importance to constrain the extragalactic background light
  (EBL). The $\gamma$-ray flux from distant blazars is in fact absorbed on
  its way from the source through its interaction with EBL photons. One can
  then use the measurement of the induced distortions to derive constraints
  on the EBL density \citep[e.g.][]{maz07}. The discovery of high redshift,
  $\gamma$-ray emitting blazars, strongly detected in the \fermi highest
  energy band ($10 - 100$ GeV) in the case of RX J0035.2+1515 and CRATES
  J0630$-$2406 (see Figs. \ref{fig:0035} and \ref{fig:0630}), is therefore
  very important also in this context.  Some of the high-redshift sources
  of the type described in this paper may be detectable by existing
  Cherenkov telescopes, as the $10 - 100$ GeV \fermi fluxes of both sources
  discussed above are within the range of observed values of blazars that
  have already been detected at TeV energies. The next generation of VHE
  telescopes, particularly the Cherenkov Telescope Array, will be able to
  detect many such sources, especially if the low energy threshold will
  reach $\sim 10$ GeV, as planned.

\end{enumerate}

\section*{Acknowledgments}
We acknowledge the use of data and software facilities from the ASI Science
Data Center (ASDC), managed by the Italian Space Agency (ASI). Part of this
work is based on archival data and on bibliographic information obtained
from the NASA/IPAC Extragalactic Database (NED) and from the Astrophysics
Data System (ADS) Bibliographic Services. Part of the funding for GROND
(both hardware as well as personnel) was generously granted from the
Leibniz-Prize to Prof. G. Hasinger (DFG grant HA 1850/28-1).

\label{lastpage}

\end{document}